\documentclass[conference]{IEEEtran}

\usepackage[utf8]{inputenc}
\usepackage[english]{babel}
\usepackage[pdftex]{graphicx}
\usepackage{amsmath}
\usepackage{verbatim}
\usepackage{url}
\usepackage[small]{subfigure}
\usepackage{multirow}
\usepackage{array}
\usepackage{slashbox}

\DeclareGraphicsExtensions{.jpeg,.pdf,.png}

\begin{document}
 
\title{Constructing Performance Models for Dense Linear Algebra Algorithms on Cray XE Systems}
 
\author{\IEEEauthorblockN{Jorge Gonz\'alez-Dom\'inguez\IEEEauthorrefmark{1}$^1$, Evangelos Georganas\IEEEauthorrefmark{2}$^1$, Yili Zheng\IEEEauthorrefmark{3}, Mar\'ia J. Mart\'in\IEEEauthorrefmark{1}}\\
\IEEEauthorblockA{\IEEEauthorrefmark{1}Computer Architecture Group, University of A Coru\~{n}a, A Coru\~{n}a, Spain\\}
\IEEEauthorblockA{\IEEEauthorrefmark{2}Dept. of Electrical Engineering and Computer Sciences, University of California at Berkeley, Berkeley, CA 94720\\}
\IEEEauthorblockA{\IEEEauthorrefmark{3}Lawrence Berkeley National Laboratory, Berkeley, CA 94720\\}}
 
%\thanks{This research was supported in part by the U.S. Government. The views and conclusions contained in this document are those of the authors and should not be interpreted as representing the official policies, either expressed or implied, of the U.S. Government.} 

\maketitle              % typeset the title of the contribution
\markboth{}{}
\pagestyle{empty} 
\thispagestyle{empty}

\begin{abstract}
Hiding or minimizing the communication cost is key in order to obtain good performance on large-scale systems. While communication overlapping attempts to hide communications cost, 2.5D communication avoiding algorithms improve performance scalability by reducing the volume of data transfers at the cost of extra memory usage. Both approaches can be used together or separately and the best choice depends on the machine, the algorithm and the problem size. Thus, the development of performance models is crucial to determine the best option for each scenario. In this paper, we present a methodology for constructing performance models for parallel numerical routines on Cray XE systems. Our models use portable benchmarks that measure computational cost and network characteristics, as well as performance degradation caused by simultaneous accesses to the network. We validate our methodology by constructing the performance models for the 
2D and 2.5D approaches, with and without overlapping, of two matrix multiplication algorithms (Cannon's and 
SUMMA), triangular solve (TRSM) and Cholesky. We 
compare the estimations provided by these models with the experimental results using up to 24,576 cores of a Cray XE6 system and predict the performance of the algorithms on larger systems. Results prove that the estimations significantly improve when taking into account network contention.
\footnotetext[1]{Jorge and Evangelos contributed equally to this work.} 

\end{abstract}
 

\section{Introduction}
\label{sec:intro}

In spite of the improvement of the interconnection networks during recent years, communication cost is one of the most significant factors in application performance and the cost of data movement within and between nodes will continue to grow relative to the cost of computation. In fact, the effective bandwidth of many current networks decreases when many messages are concurrently sent from different cores.

Two techniques to minimize the impact of communication cost are {\it communication avoidance} and {\it communication overlapping}. The first one reduces the volume of communications~\cite{SoDe11,SoDe12} while the latter hides communication impact on the application performance by overlapping communications with computational work. Nevertheless, complicated interactions and trade-offs arise when they are used together. Our previous work~\cite{GeGo12} studied how to combine and balance these two techniques for various linear algebra algorithms. Specifically, these techniques were applied to two matrix multiplication algorithms (Cannon's and SUMMA), triangular solve (TRSM) and Cholesky factorization. We demonstrated that the combination of both techniques is usually effective and significantly reduces execution time in most cases. However, depending on the algorithm, the problem size and the number of cores, using only overlapping can be more efficient. 

The goal of this work is to create a methodology for constructing performance models that predict the effects of communication avoidance and overlapping on Cray XE systems. An important application of these performance models would be in guiding performance optimizations such as selecting the appropriate technique and tuning parameters. The new methodology includes a calibration factor to take into account performance degradation when concurrent communications from multiple nodes occur. It considers the performance of numerical computations, the latency and the ideal communication bandwidth of the network, as well as this calibration factor. These data are measured through completely portable benchmarks, so this methodology can be used for other architectures. The performance models obtained through the new methodology were evaluated on four algorithms (Cannon's, SUMMA, TRSM and Cholesky 
factorization), although they are easily adapted to other numerical algorithms. As shown by the experimental results, the inclusion of the calibration factor makes model predictions much more accurate.

The rest of this paper is organized as follows. Section~\ref{sec:relatedWork} summarizes the related work. Section~\ref{sec:platform} presents the platform used for the experiments. Section~\ref{sec:methodology} describes the methodology followed to develop the models and the benchmarks employed. In Section~\ref{sec:models} we explain in detail the models for two examples, the Cannon's and TRSM algorithms. Section~\ref{sec:results} compares the estimations of the models with the experimental results on Hopper, a Cray XE6 system, and gives performance predictions for our algorithms on larger systems. Finally, conclusions are discussed in Section~\ref{sec:conclusions}.

\section{Related Work}
\label{sec:relatedWork}

Communication avoidance and overlapping are techniques used to improve application performance on large supercomputers. Overlapping communication with computation or with other communication can increase efficiency without decreasing the communication volume. The benefits of overlapping can be very significant, for example, in one-sided communication~\cite{BeBo06,NiHa09} and in collective communication~\cite{NiZh11}. On the other hand, the so-called 2.5D algorithms take advantage of extra memory to reduce the volume of communication along the critical path. The motivation of this type of communication avoiding algorithms and the explanation of why they attain the communication lower bounds can be found in~\cite{SoDe11} and~\cite{SoDe12}. These works also present the MPI implementation of some 2.5D algorithms and their experimental evaluation. As indicated in Section~\ref{sec:intro}, the interaction of both techniques for UPC-based implementations of Cannon's, SUMMA, TRSM and Cholesky algorithms 
was thoroughly evaluated in~\cite{GeGo12}.

Regarding performance modeling, a well known communication model for distributed memory is the LogP family~\cite{CuKa93, AlIo97, YuZh10}. However, these models do not take into account the bandwidth degradation caused by the sharing of resources in large-scale systems, which has been proved as a crucial factor for communication performance on modern systems~\cite{HoJo06}. Several recent works began to take this issue into account to model current parallel applications. Hoefler and Snir~\cite{HoSn11} provided a definition of the terms {\it dilation} and {\it congestion} and considered them to generate efficient mapping policies for MPI applications. Bhathel\'e and Kal\'e~\cite{BhLa09} studied the effects of contention on large supercomputers with torus and mesh network topologies. They conclude that, in presence of contention, message latencies increase significantly with the distance between the communicated nodes. Performance models for interconnection networks that take congestion into 
consideration 
were presented in~\cite{JeMa08} and~\cite{MaMe11}. These models measure degradation of bandwidth using graphs and need a deep knowledge of the network schemes of communication. 

Our model is based on benchmarking and can be considered as an adaptation to current systems of the $\alpha-\beta$ model used in~\cite{Ho94} to estimate communication times. It takes into account the features of the modern machines by including factors that incorporate communication performance degradations.

There is also research done for shared memory platforms that consider contention as a key factor. For instance, Gibbons et al.~\cite{GiMa97} take into account memory contention while Helman and J\'aj\'a~\cite{HeJa01} also include processor contention. A two-level model that characterizes the impact of contention and cache effects and, at the same time, develops and studies a graph model of the application is presented in~\cite{AdVe04}. There exist similar works for hybrid MPI/OpenMP computation, that can help to decide the best combination of MPI processes and OpenMP threads depending on the characteristics of the machine and the code~\cite{AdCh07,LiGo08,WuTa11}.

\section{Experimental Platform}
\label{sec:platform}

Before the explanation of the performance models, we describe the hardware and software environments used in the experiments because they are helpful to understand some of the results of the benchmarks.

Our target system is Hopper, a Cray XE6 supercomputer with 153,216 compute cores and 217 TB of memory in total. Each Cray XE6 node has 24 cores, grouped by 6 in 4 Non-Uniform Memory Access (NUMA) domains. CPU cores have faster access speed to  memory within the same NUMA domain. Inter-node communication is done through the custom Cray Gemini Network, which is a high-bandwidth and low-latency 3D torus interconnect with hardware RDMA support. Table~\ref{tab:platform} lists the specifications of our experimental system.

\begin{table}[!ht]
  \centering
  \begin{tabular}{|l|l|}
    \hline 
    System & Cray XE6 (Hopper) \\ 
    \hline 
    Processor & AMD Opteron ``Magny-Cours'' \\ 
    Clock rate & 2.1~GHz \\ 
    Peak performance per core & 8.4~Gflops \\ 
    Cores per NUMA domain & 6 \\ 
    NUMA domains per node & 4 (packaged in 2 sockets) \\ 
    Total cores per node & 24 \\ 
    Private L1 data cache per core & 64~KB \\ 
    Private L2 data cache per core & 512~KB \\ 
    Shared L3 cache per NUMA domain & 6~MB \\ 
    Memory bandwidth & 25.6 GB/s \\ 
    Memory per node & 32 GB DDR3-1066 ECC \\ 
    Compiler & Cray Compiler \\ 
    Interconnect & Gemini 3D Torus \\ 
    Peak Bandwidth (per direction) & ~7 GB/s \\ 
    \hline
  \end{tabular}
  \caption{Specifications of the Hopper supercomputer}
  \label{tab:platform}
\end{table}

The parallel implementations of the Cannon's, SUMMA, TRSM and Cholesky factorization algorithms presented in~\cite{GeGo12} are used as benchmark applications to test the methodology for constructing the proposed performance models. Although the codes are mainly written using UPC, the implementations employ three different parallel programming models (UPC, MPI and Pthreads) to exploit the hardware potential. These parallel algorithms use local linear algebra routines provided by the LibSci library, a collection of multithreaded numerical routines optimized for best performance on Cray systems. Thus, a hybrid (process/thread) parallelization model is used. Specifically, we run one UPC process per NUMA domain and one thread per core within the NUMA region. Each UPC process is mapped into an OS process and each process uses 6 OS threads, 1 thread per core. MPI collectives are used to supplement the UPC implementations in order to overcome some limitations of the current UPC collective library~\cite{GeGo12}.

\section{Description of the Methodology}
\label{sec:methodology}

In this section we describe the general methodology to develop detailed performance models for linear algebra algorithms. They are constructed by using a divide-and-conquer methodology, which tracks the execution flow of each algorithm and estimates the completion time for every encountered operation. These operations are identified as computation or communication. In regard to the overlapped operations, the models predict the execution time as the maximum expected completion time of each individual operation. System parameters of the target computer extracted through portable benchmarks are used for the estimations.

Computation operations refer to the numerical computations that are performed by each process using its local data. Parallel algorithms usually employ multithreaded linear algebra routines provided by vendor optimized libraries for local computation. Thus, their completion time depends on the computational efficiency of the routine on the system, the number of underlying threads and the problem size. Micro-benchmarks are executed on the target platform to obtain the efficiency of each employed routine. Functions $T_{rout}(d, t)$ return within the models the time to perform the computations of matrices with size $d$ calling the $rout$ routine with $t$ underlying threads. Both the benchmarks and the estimation functions are developed to predict the computational time of problems that work with square matrices. Operations with rectangular matrices are estimated as several consecutive square matrix operations.

\begin{figure}[!ht]
  \centering
  \includegraphics[width=0.48\textwidth]{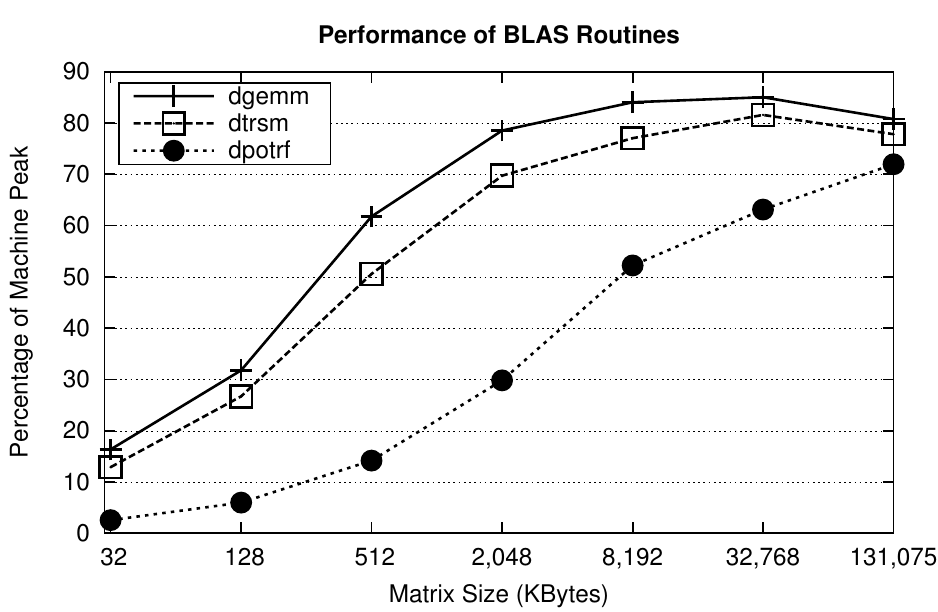}
  \caption{Efficiency of several LibSci BLAS routines on Hopper. Run with 6 cores in a NUMA domain.}
  \label{fig:BLAS_perf}
\end{figure}

Figure~\ref{fig:BLAS_perf} illustrates the efficiency of the BLAS routines used by the benchmark applications on Hopper for different matrix sizes. As in the parallel benchmarks, six threads (one per core within the NUMA domain) were used to run the LibSci routines.

The second type of operations is communication. In an ideal scenario, the time to transfer data between two processors should be easily determined by only knowing the network latency, the contention-free bandwidth between two nodes and the message size. If $w$ the number of transferred elements, $L$ the latency of the system, and $\beta$ the inverse bandwidth (in seconds/word), we estimate itassume as:
\setlength\arraycolsep{0.1em}
\begin{small}
\begin{eqnarray}\nonumber
 T_{comm\_ideal}(w) &=& L+\beta *w
\end{eqnarray}
\end{small}

\begin{figure}[!ht]
  \centering
  \includegraphics[width=0.48\textwidth]{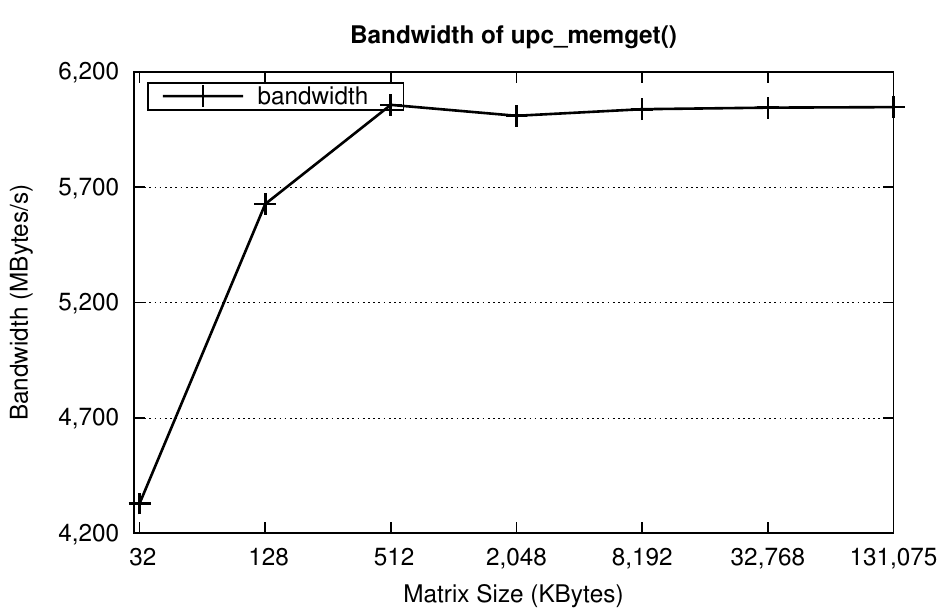}
  \caption{Inter-node contention-free network bandwidth on Hopper using UPC one-sided communication.}
  \label{fig:BandWidth}
\end{figure}

Latency and bandwidth are obtained through the benchmarks presented for the LogP model~\cite{CuKa93}. It uses only two processes placed on different nodes and measures the communication times for different message sizes. Several executions are performed for each message size and the average communication times are used to calculate latency and bandwidth. Figure~\ref{fig:BandWidth} shows the contention-free bandwidth of Hopper's network when UPC one-sided communication is used.

However, sharing the same network links among several nodes leads to a significant communication performance degradation in real environments where several parallel processes (created by the same or other programs) are transferring data simultaneously. A new micro-benchmark is therefore included to determine what we call ``calibration factor'', that is, the rate between ideal and real communication costs when several processes use the network simultaneously. The structure of this microbenchmark is the same as for the calculation of the ideal bandwidth in the LogP model but, in this case, several processes send the same message at the same time. Only one process per NUMA region is considered as multithreaded libraries are usually available and its use is recommended for the parallelization of linear algebra algorithms. The microbenchmark is executed 
varying 
the message size, the number of processes considered and the ``communication distance'' between processes. We define ``communication distance'' as the rank difference between the source process and the destination process. Although the mapping of processes to nodes is independent of the rank (processes with close ranks do not need to be scheduled on close nodes), generally larger distances imply messages moving along more links of the network. %We assume the same communication distance from all processes, which is quite common for parallel linear algebra algorithms.

\begin{figure}[ht]
  \centering
  \includegraphics[width=0.48\textwidth]{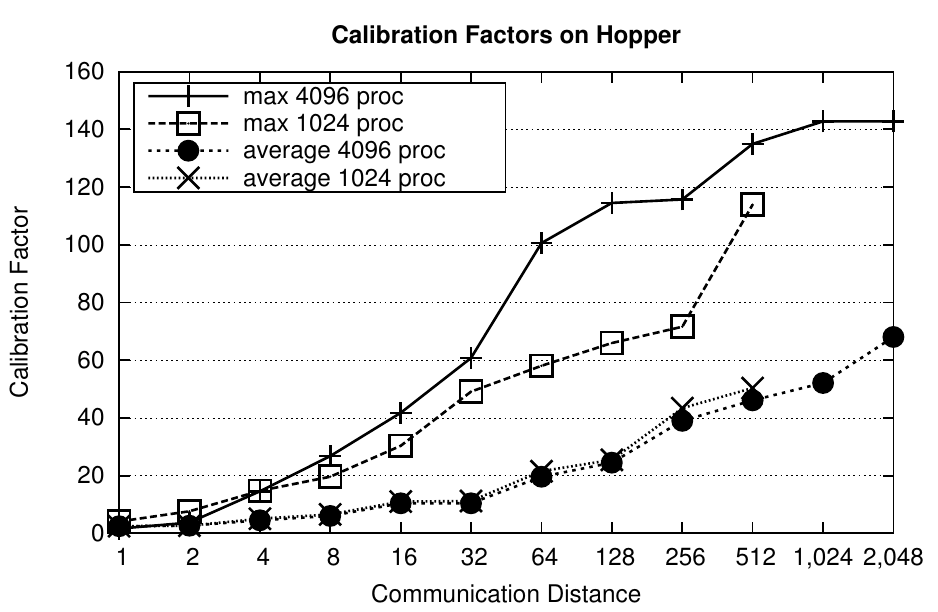}
  \caption{Calibration factors on Hopper depending on the distance of the processes that exchange the data.}
  \label{fig:congestion}
\end{figure}

For instance, Figure~\ref{fig:comm_threads} shows the time needed by each process to perform 10 communication operations of 64MB, with a communication distance of 16 (64 processes were used for this experiment) and all the processes carrying out the communication at the same time. As can be seen in the figure, there is a huge variance among the times obtained by different processes. Degradation is not regular in all processes. Therefore, two calibration factors are extracted from each one of the micro-benchmark executions: the average and the maximum. Figure~\ref{fig:congestion} shows these two calibration factors for 1,024 processes (6,144 cores) and 4,096 processes (24,576 cores) on Hopper. Messages of size 64MBytes were employed for this figure but additional experiments were performed changing the message size. The following conclusions were derived from these experiments:

\begin{figure}[ht]
  \centering
  \includegraphics[width=0.48\textwidth]{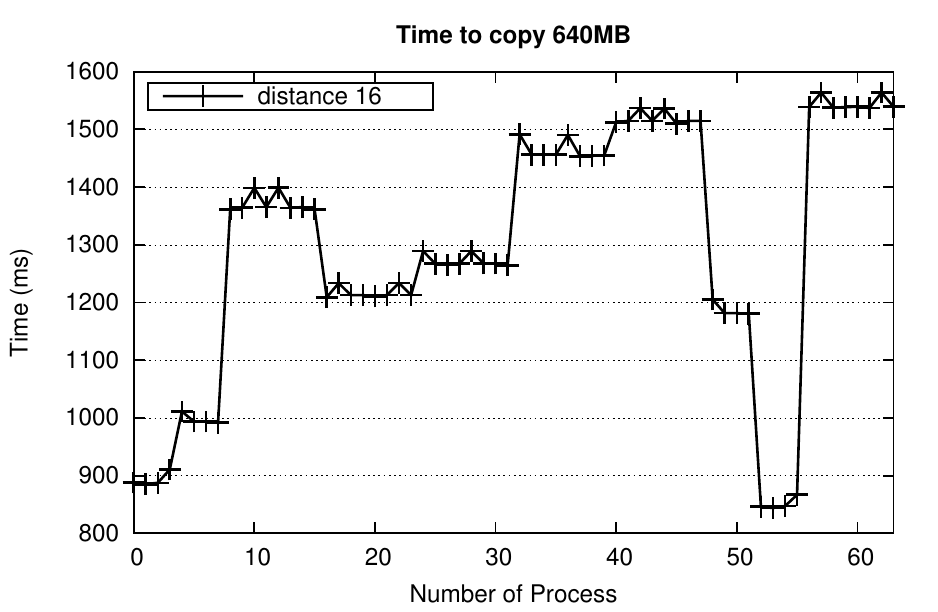}
  \caption{Completion time of each process to transfer 640MB using ten communications with a communication distance of 16 on Hopper (64 processes perform the same operation at the same time).}
  \label{fig:comm_threads}
\end{figure}

\begin{itemize}
 \item Both calibration factors do not significantly depend on the message size if messages above 256KBytes are used.
 \item The average calibration factor does not significantly depend on the total number of processes used at the same time.
  \item The maximum calibration factor significantly depends on the total number of processes used at the same time. 
 \item Both the average and the maximum calibration factors significantly depend on the communication distance. For instance, whatever the total number of processes is, the factor when all processes transfer data with a communication distance of 32 is much higher than the factor with a communication distance of 4. The increase of latency due to message distance on modern large-scale supercomputers with 3D networks had been previously analyzed in~\cite{BhLa09}.
\end{itemize}

Therefore, the average calibration factor only depends on the communication pattern while the maximum also depends on the total number of processes performing a communication at the same time. Two functions were created in order to include these factors in the models:

\begin{itemize}
 \item $C_{avg}(d)$: It returns the average calibration factor when all the participating processes perform communication at the same time with a communication distance $d$.\\
 
 \item $C_{max}(p, d)$: It provides the maximum calibration factor when $p$ processes perform communication at the same time with a communication distance $d$.\\
\end{itemize}

In order to estimate the communication cost, we start executing the microbenchmark to determine the impact of network contention using a representative message size of 64MB and different numbers of processes and communication distances. Then, we obtain the calibration factors as the ratio between the ideal and the real communication time obtained on each scenario. For each number of threads we use a different execution on order to not leave nodes idle. These factors will be used into the models to better estimate the communication times. In presence of synchronizations all threads have to wait for the one with the highest communication cost, thus the maximum calibration factor will be used when synchronization among processes is required and the average will be used otherwise. Being $p$ the total number of processes and $d$ the communication distance, the completion time of each communication operation is estimated as follows (without and with synchronization, respectively):

\begin{small}
\begin{eqnarray}\nonumber
 T_{comm}(w, d) &=& C_{avg}(d)*(L+\beta *w)
\end{eqnarray}
\begin{eqnarray}\nonumber
 T_{comm\_sync}(p, w, d) &=& C_{max}(p, d)*(L+\beta *w)
\end{eqnarray}
\end{small}

\section{Performance Models}
\label{sec:models}

The methodology proposed in the previous section was applied to model the implementations of Cannon's, SUMMA, TRSM and Cholesky factorization presented in~\cite{GeGo12}. In order to simplify the paper, only the Cannon's and TRSM algorithms are discussed in detail as representative examples.

\subsection{Cannon's Algorithm}
\label{sec:cannon}

The traditional 2D Cannon's algorithm~\cite{Ca69} performs the matrix-matrix multiplication ($A*B=C$) by shifting blocks of data among near neighbor processes on a 2D grid of $p$ processes, with square blocks of size $n/\sqrt{p}$ (being $n$ the total matrix size). The algorithm starts by performing a skew on the initial matrices along rows and columns of the process grid. The blocks are lined up so that at each subsequent step only a single shift needs to be done between each block multiplication. Cannon's algorithm uses near-neighbor point-to-point remote copies rather than collective communications.

%\subsubsection{2D Cannon's Performance Models}
%\label{sec:cannon_model}

The 2D Cannon's algorithm presents the following structure:

\begin{enumerate}
 \item An initial shift where each process copies a block of matrix $A$ from one process that is in its row of the grid and one block of matrix $B$ from one process that is in its column of the grid.
 \item A loop with $\sqrt{p}-1$ iterations where:
 \begin{enumerate}
  \item Each process calls {\it dgemm} with the current blocks.
  \item The processes in the same rows and in the same columns of the grid are synchronized to guarantee that all of them have received the blocks that will be copied in the next iteration.
  \item Each process copies the next blocks. The block of $A$ is copied from the next process in the row ($MYTHREAD$\footnote[2]{$MYTHREAD$ is the UPC indentifier for the thread that is performing that part of the code}$+1$) and the block of $B$ from the next process in the column ($MYTHREAD+\sqrt{p}$).
 \end{enumerate}
 \item A {\it dgemm} operation with the final blocks. 
\end{enumerate}

As explained in Section~\ref{sec:methodology} the estimation of the execution time for the whole algorithm is the addition of the estimation of all parts. Thus, it can be specified as:
\begin{small}
\begin{eqnarray}\nonumber
 T_{Cannon\_2D}&=&T_{iniShiftRow}+T_{iniShiftCol}+(\sqrt{p}-1)*\\\nonumber
 &&\Big(T_{shiftRow}+T_{shiftCol}+T_{multBlock}\Big)+\\\nonumber
 &&T_{multFinalBlock}
\end{eqnarray}
\end{small}

This equation is easily simplified because the initial shifts and the final {\it dgemm} are equal to the shifts and products of the loop, respectively. As explained in Section~\ref{sec:methodology}, $T_{dgemm}$, obtained from the execution of the multithreaded $dgemm$ BLAS routine on the target platform, provides an estimation of the computational time of this routine. Regarding the communications, some considerations must be taken into account: 

\begin{itemize}
 \item Shifts among processes in the same column of the grid involve a communication of distance $\sqrt{p}$, whereas shifts among processes in the same row of the grid involve a communication of distance 1.
 \item Due to the synchronizations within the loop, all processes must wait for the last process that finishes the shifts. The effect of this synchronization is included in the model by using the maximum calibration factor.
\end{itemize}

With all these assumptions the completion time of each communication operation (including the impact of the synchronization) is estimated using $T_{comm\_sync}$ (see Section~\ref{sec:methodology}). Consequently, as the communication distances in the shifts by rows and columns are, respectively, 1 and $\sqrt{p}$, the model for the 2D Cannon's algorithm is:
\begin{small}
\begin{eqnarray}\nonumber
T_{Cannon\_2D} &=& \sqrt{p}*\Big(T_{comm\_sync}(p, bs^{2}, 1)+\\\nonumber
&&T_{comm\_sync}(p, bs^{2}, \sqrt{p})+T_{dgemm}(bs,t)\Big)
\end{eqnarray}
\end{small}

\noindent $bs$ being the block size ($n/\sqrt{p}$) and $t$ the number of underlying threads used to run the local numerical routine.

Overlapping is included in this algorithm by moving forward the shifts of the next iteration of the loop and hiding their cost by using asynchronous communications while performing the products with the blocks of the current iteration. We consider that the algorithm achieves perfect overlapping, so the completion time is estimated as the maximum of the two operations. The model is the addition of the completion time of the first shift, the final {\it dgemm} (which cannot be overlapped) and the overlapped loop:  
\begin{small}
\begin{eqnarray}\nonumber
T_{Cannon\_2D\_ovlp} &=& T_{comm\_sync}(p, bs^{2},1)+\\\nonumber
&&T_{comm\_sync}(p, bs^{2},\sqrt{p})+\\\nonumber
&&T_{dgemm}(bs,t)+(\sqrt{p}-1)*\\\nonumber
&&max\Big[T_{comm\_sync}(p, bs^{2},1)+\\\nonumber
&&T_{comm\_sync}(p, bs^{2},\sqrt{p}), T_{dgemm}(bs,t)\Big]
\end{eqnarray}
\end{small}

%\subsubsection{2.5D Cannon's Performance Models}
The 2.5D algorithm initially replicates the blocks of $A$ and $B$ on $c$ layers and redundantly performs independent updates on the copies of $C$, which are combined using a reduction operation at the end. The blocks size is $n/(\sqrt{p}/c)$ but there are only $\sqrt{p}/c$ shifts, decreasing the total amount of communications. The initial replications are performed using communications whose source process is always on the first layer. The communication distance depends on the target layer. The worst case (distance to the last layer) will be used for the estimation:
\begin{small}
\begin{eqnarray}\nonumber
 T_{iniRepl}(p, w, c) &=& 2*C_{max}(p, (c-1)*p/c)*(L+\beta *w)
\end{eqnarray}
\end{small}

% \begin{figure}[!ht]
%   \centering
%   \includegraphics[width=0.4\textwidth]{figures/Reduce.pdf}
%   \caption{Reduction operation with binomial tree.}
%   \label{fig:Reduce}
% \end{figure}

The final reduction is performed using the MPI collective reduce operation. The algorithms used for collective communications depend on the MPI implementation. We assume that the behavior of collective operations is the one specified in~\cite{ThRa05} where the authors study different algorithms to perform these operations and indicate which is the best option depending on the number of processes and the message size. Taking into account that power-of-two numbers of processes and medium-sized messages are used by our benchmark applications, the Rabenseifner's algorithm~\cite{Ro04}, a reduce-scatter followed by a gather to the root with a synchronization between them, is assumed for the reduce operation. Considering the parameter $d$ as the closest communication distance between two of the processes involved in the reduce, $q$ the number of processes involved in the reduction, $p$ the total number of processes, and $w$ the size of the block to reduce, the estimation function is:

\begin{small}
\begin{eqnarray}\nonumber
 T_{reduce}(p, q, w, d) &=& T_{redSca\_sync}(p, q, w, d)+T_{gather}(q, w, d)
\end{eqnarray}
\end{small}

The reduce-scatter follows the recursive halving algorithm (see~\cite{ThRa05}). If there are $q$ processes involved in the reduction, there are $\log_2(q)$ steps where each process exchanges data with a process that is in a distance of $2^{i}$ ($i$ being the number of the step). Consequently, each step increases the communication distance. In order to include the impact of the synchronization between the reduce-scatter and the gather, the last step uses the maximum calibration factor. For the rest of the steps the average factor is used:
\begin{small}
\begin{eqnarray}\nonumber
 T_{redSca\_sync}(p, q, w, d) &=& \sum_{i=0}^{\log_2(q)-2} \Big(C_{avg}(2^{i}*d)*\\\nonumber
 &&(L+\beta *w*q/2^{i})\Big)+\\\nonumber
 &&C_{max}(p, 2^{\log_2(q)-1}*d)*\\\nonumber
 &&(L+\beta *w*t/2^{\log_2(q)-1})
\end{eqnarray}
\end{small}

The binomial tree algorithm is used for the gather. The average calibration factor is always employed in the equation as there is no synchronization at the end:
\begin{small}
\begin{eqnarray}\nonumber
 T_{gather}(q, w, d) &=& \sum_{i=0}^{\log_2(q)-1} \Big(C_{avg}(2^{i}*d)*(L+\beta *(w/q)*2^{i})\Big)
\end{eqnarray}
\end{small}

Therefore, knowing that $c$ is the number of layers and thus the block size is $n/\sqrt{p/c}$, the performance models for the 2.5D algorithms are:
\begin{small}
\begin{eqnarray}\nonumber
T_{Cannon\_2.5D} &=& T_{iniRepl}(p,bs^{2},c)+(\sqrt{p/c}-1)*\\\nonumber
&&\Big(T_{comm}(p,bs^{2},1)+T_{comm}(p,bs^{2},\sqrt{p/c}) +\\\nonumber
&&T_{dgemm}(bs,t)\Big) +T_{dgemm}(bs,t) +\\\nonumber
&&T_{reduce}(p, c, bs^{2}, p/c)
\end{eqnarray}
\begin{eqnarray}\nonumber
T_{Cannon\_2.5D\_ovlp} &=& T_{iniRepl}(p,bs^{2},c)+(\sqrt{p/c}-1)*\\\nonumber
&&max\Big[T_{comm}(p,bs^{2},1)+\\\nonumber 
&&T_{comm}(p,bs^{2},\sqrt{p/c}),T_{dgemm}(bs,t)\Big] +\\\nonumber
&&T_{dgemm}(bs,t) +T_{reduce}(p, c, bs^{2}, p/c)
\end{eqnarray}
\end{small}

\subsection{Triangular Solve}
\label{sec:trsm}

Triangular solve (TRSM) is used to compute a matrix $X$, such that $X*U = B$, where $U$ is upper-triangular and $B$ is a dense matrix. This problem has dependencies across the columns of X, while each row of X can be computed independently. 
%For $n$-by-$n$ matrices, solving this problem has the same asymptotic computational cost as matrix multiplication. The problem has also been shown to require at least as much communication as matrix multiplication~\cite{BaDe11}.

%\subsubsection{2D Triangular Solve Performance Models}
%\label{sec:trsm_model}

The 2D algorithm is based on the block version where the matrices are partitioned in blocks as:
$$\begin{bmatrix} X_{00} & X_{01} \\ X_{10} & X_{11}  \end{bmatrix} *
\begin{bmatrix} U_{00} & U_{01} \\ 0 & U_{11} \end{bmatrix} =
\begin{bmatrix} B_{00} & B_{01} \\ B_{10} & B_{11} \end{bmatrix}$$
and, at each step, a block-column of $X$ is computed. This block-column can then be used to update the trailing matrix $B$. The computation proceeds as follows:

\begin{enumerate}
\item Broadcast $U_{00}$ and $U_{01}$ along the columns of the grid.
\item Compute via {\it dtrsm} $X_{00} = B_{00}*U_{00}^{-1}$ and $X_{10} = B_{10}*U_{00}^{-1}$.
\item Once the previous {\it dtrsm} are completed, broadcast $X_{00}$ and $X_{10}$ along the rows of the grid.
\item Update via {\it dgemm} $B_{01} = B_{01} - X_{00}*U_{01}$ and $B_{11} = B_{11} - X_{10}*U_{01}$.
\item Broadcast $U_{11}$ along the columns of the grid.
\item Compute via {\it dtrsm} $X_{01} = B_{01}*U_{11}^{-1}$ and $X_{11} = B_{11}*U_{11}^{-1}$.
\end{enumerate}

Communication is minimized by employing a block-cyclic layout where each process owns multiple sub-blocks of the matrices. The number of blocks per process is indicated by the parameter $r$. Consequently,  the matrices are divided in $r*\sqrt{p}$-by-$r*\sqrt{p}$ blocks and the block size is $bs = n/(r*\sqrt{p})$. Thus, steps 1-4 are computed inside a loop of $r*\sqrt{p}$ iterations.

The main part of the algorithm is a loop with four operations per iteration. The estimation is the addition of the completion time of each iteration plus the final broadcast and {\it dtrsm}:
\begin{small}
\begin{eqnarray}\nonumber
 T_{TRSM\_2D} &=& \sum_{i=0}^{r*\sqrt{p}-1} \Big(T_{bcast\_U} + T_{solve\_iter} + T_{bcast\_X} +\\\nonumber
 &&T_{update}\Big) + T_{last\_bcast\_U} + T_{last\_solve}
\end{eqnarray}
\end{small}

The main difference with the matrix-matrix multiplication algorithms is that the workload changes according to the iteration. The more iterations pass, the less blocks of $U$ are involved in the update and the broadcast along the columns of the grid.
Besides, most of computational operations work with rectangular matrices which are estimated as several calls using square matrices, as explained in Section~\ref{sec:methodology}.

The broadcasts and gathers in the different TRSM versions are performed employing MPI. Following again the work~\cite{ThRa05}, the MPI broadcast is assumed to be performed through a scatter followed by an all-gather (with a synchronization between them):
\begin{small}
\begin{eqnarray}\nonumber
 T_{bcast}(p, q, w, d) &=& T_{scatter\_sync}(p, q, w, d) +T_{all-gather}(q, w, d)
\end{eqnarray}
\end{small}

The models for $T_{scatter\_sync}$ and $T_{all-gather}$ are exactly the same as the ones for $T_{redSca\_sync}$ and $T_{gather}$ presented for Cannon's algorithm, respectively. Finally, it must be taken into account that there is one synchronization per iteration: the broadcast of $X$ along rows cannot start until the previous computations have finished. This synchronization is modeled by using $T_{bcast\_sync}$ for the broadcast of $U$. It uses the maximum calibration factor in the last of the $\log_2(q)$ steps of the gather. These assumptions lead to the following model:
\begin{small}
\begin{eqnarray}\nonumber
 T_{TRSM\_2D} &=& \sum_{i=0}^{r*\sqrt{p}-1} \Big(((r*\sqrt{p}-i)/\sqrt{p})*\\\nonumber
 &&T_{bcast\_sync}(p,\sqrt{p},bs^{2},\sqrt{p})+r*(T_{dtrsm}(bs, t)+\\\nonumber
 && T_{bcast}(p, \sqrt{p},bs^{2},1))+((r*\sqrt{p}-i-1)/\sqrt{p})*\\\nonumber
 &&T_{dgemm}(bs^{2},t))\Big)+r*T_{dtrsm}(bs, t)+\\\nonumber
 && T_{bcast\_sync}(p,\sqrt{p},bs^{2},\sqrt{p})
 \end{eqnarray}
 \end{small}

The algorithm is optimized by overlapping the broadcast of the next iteration of $U$ with the matrix update. In this case, Pthreads are used so that one of the threads is completely dedicated to communications. Therefore, these communications are overlapped with the numerical computation performed by the other $t-1$ threads.

\begin{small}
\begin{eqnarray}\nonumber
  T_{TRSM\_2D\_ovl} &=& r*T_{bcast\_sync}(p,\sqrt{p},bs^{2},\sqrt{p})+\\\nonumber
  &&\sum_{i=0}^{r*\sqrt{p}-1} \Big(r*(T_{dtrsm}(bs, t-1)+\\\nonumber
  && T_{bcst}(p, \sqrt{p},bs^{2},1))+((r*\sqrt{p}-i-1)/\sqrt{p})*\\\nonumber
  && max[T_{bcast\_sync}(p,\sqrt{p},bs^{2},\sqrt{p}),\\\nonumber
  && r*T_{dgem}(bs^{2},t-1)]\Big)+r*T_{dtrsm}(bs, t-1)
  \end{eqnarray}
  \end{small}

%\subsubsection{2.5D Triangular Solve Performance Models}  
  
For the 2.5D approach, the initial distribution of the matrices in one layer is the same as for the 2D algorithm: a block-cyclic distribution along the $p/c$ processes in each dimension. Thus, the block size is $n/(r*\sqrt{p/c})$-by-$n/(r*\sqrt{p/c})$. We initially replicate the triangular matrix $U$ along layers (as the input matrices in the 2.5D algorithm for the matrix product) but distribute the rows of each block of $X$ among them using one scatter operation. Remark that there are two levels of blocking for $X$, thus each process has $r^{2}$ rectangular blocks of $n/(c*r*\sqrt{p/c})$-by-$n/(r*\sqrt{p/c})$ elements. With this distribution, each layer computes a subset of the rows of $X$ via a 2D TRSM with its $\sqrt{p/c}$-by-$\sqrt{p/c}$ grid of processes. The distributed rows must be gathered once the layers finish their computation. The models for this 2.5D algorithm are:

\begin{small}
\begin{eqnarray}\nonumber
 T_{TRSM\_2.5D} &=& r^2*\Big((3/4)*T_{bcast}(p,c,bs^{2},p/c)+\\\nonumber
 &&T_{scatter\_sync}(p,c,bs^{2}/c,p/c)\Big)+\\\nonumber
 &&\sum_{i=0}^{r*\sqrt{p/c}-1} \Big(((r*\sqrt{p/c}-i)/\sqrt{p/c})*\\\nonumber
 &&T_{bcast\_sync}(p,\sqrt{p},bs^{2},\sqrt{p/c})+(r/c)*\\\nonumber
 &&(T_{dtrsm}(bs, t)+T_{bcast}(p,\sqrt{p/c},bs^{2},1)+\\\nonumber
 && ((r*\sqrt{p/c}-i-1)/\sqrt{p/c})*\\\nonumber
 &&T_{dgemm}(bs^{2},t))\Big)+\\\nonumber
 && T_{bcast\_sync}(p,\sqrt{p/c},bs^{2},\sqrt{p/c}) + (r/c)*\\\nonumber
 &&T_{dtrsm}(bs, t)+r^2*T_{gather}(c,bs^{2},p/c)
 \end{eqnarray}

 \begin{eqnarray}\nonumber
 T_{TRSM\_2.5D\_ovlp} &=& r^2*\Big((3/4)*T_{bcast}(p,c,bs^{2},p/c)+\\\nonumber
 &&T_{scatter\_sync}(p,c,bs^{2}/c,p/c)\Big)+\\\nonumber
 && r*T_{bcast\_sync}(p,\sqrt{p/c},bs^{2},\sqrt{p/c})+\\\nonumber
 && \sum_{i=0}^{r*\sqrt{p/c}-1} \Big((r/c)*(T_{dtrsm}(bs, t-1)+\\\nonumber
 && T_{bcast}(p,\sqrt{p/c},bs^{2},1))+\\\nonumber
 &&((r*\sqrt{p/c}-i-1)/\sqrt{p/c})*\\\nonumber
 && max[T_{bcast\_sync}(p,\sqrt{p/c},bs^{2},\sqrt{p/c}),\\\nonumber
 &&(r/c)*T_{dgemm}(bs^{2},t-1)]\Big)+\\\nonumber
 && (r/c)*T_{dtrsm}(bs, t-1)+\\\nonumber
 &&r^2*T_{gather}(c,bs^{2}, p/c)
 \end{eqnarray}
\end{small}

\section{Experimental Evaluation}
\label{sec:results}

This section evaluates the performance models for Cannon's, SUMMA, TRSM and Cholesky factorization using the methodology explained in this work. Hopper, the supercomputer presented in Section~\ref{sec:platform}, was used as testbed. We study not only the accuracy of the estimations provided by the models compared with the real performance, but also the predictions that they provide for scenarios with more resources.

\subsection{Validation of the Proposed Models}

Experimental results of the benchmarking algorithms were obtained on Hopper in order to compare them with the estimations provided by the models. The graphs presented in this section use the percentage of peak flops of the target system as performance measure. The comparison is shown for 6,144 and 24,576 cores (1,024 and 4,096 processes). In order to illustrate that the inclusion of the calibration factor is key to obtain good estimations both results with and without considering the calibration factor are included in the figure ({\tt est\_Cal} and {\tt est\_NoCal}, respectively).

\begin{figure*}[!ht]
  \centering
  \includegraphics[width=0.48\textwidth]{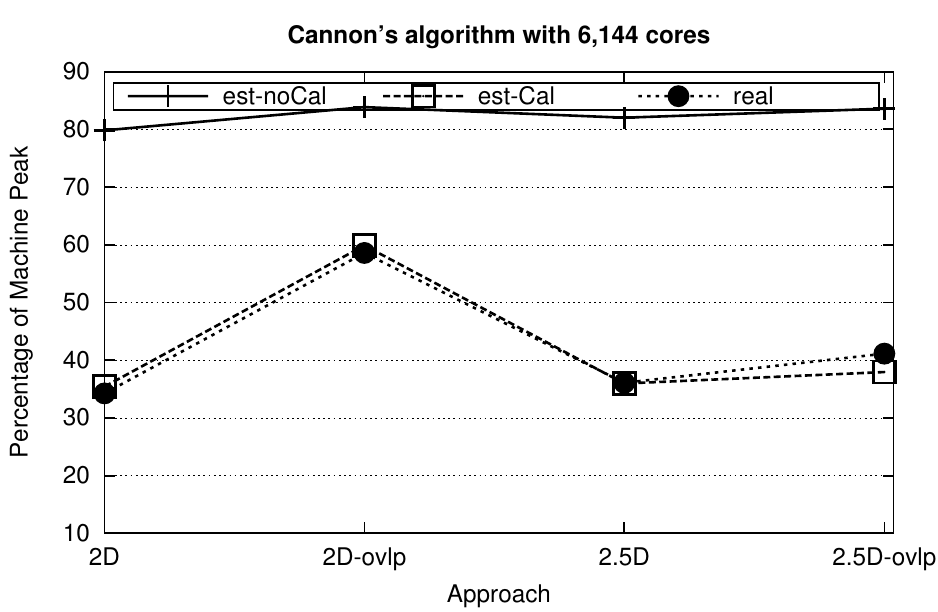}
  \includegraphics[width=0.48\textwidth]{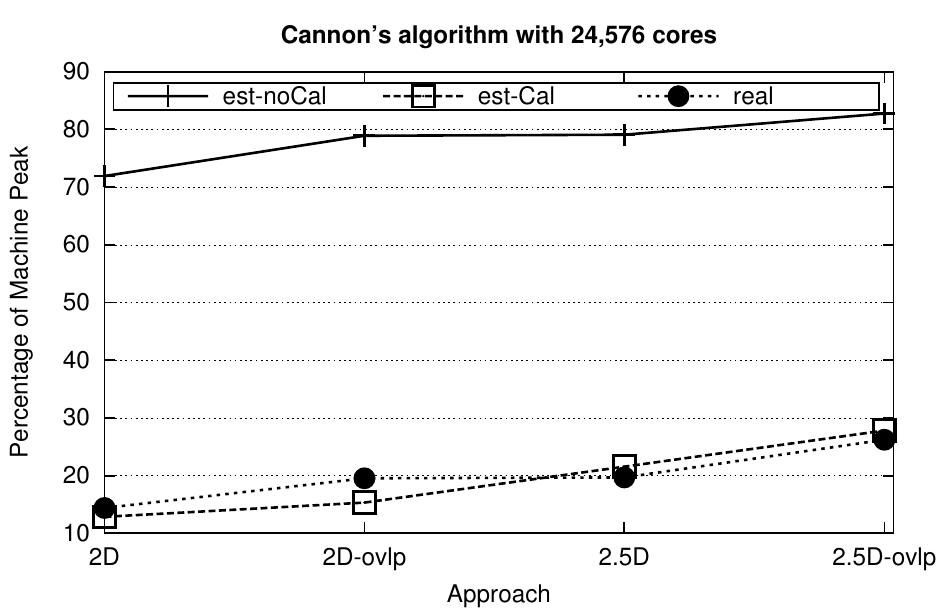}
  \caption{Real and estimated performance for Cannon's using matrices of 32,768$\times$32,768 doubles}
  \label{fig:cannon}
\end{figure*}

\begin{figure*}[!ht]
  \centering
  \includegraphics[width=0.48\textwidth]{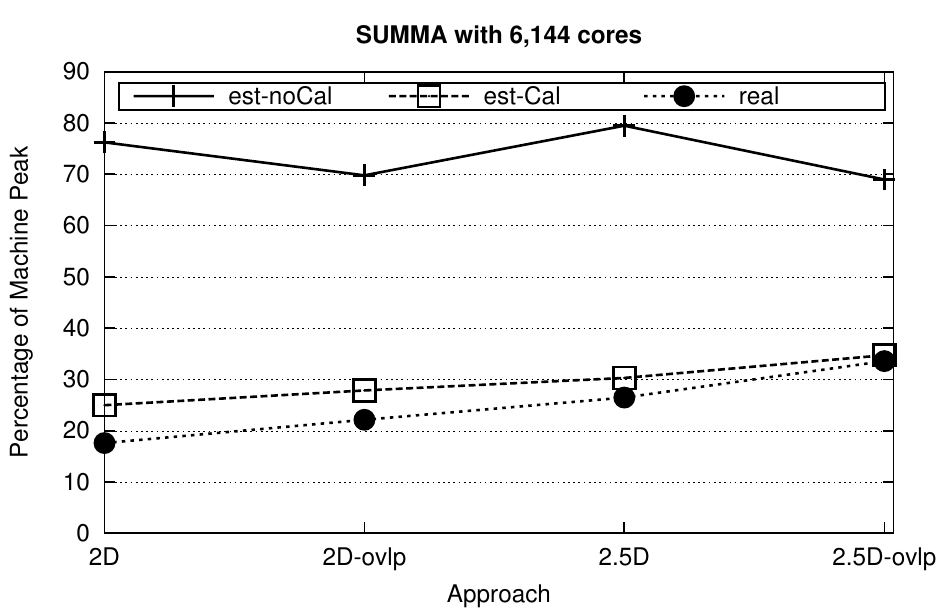}
  \includegraphics[width=0.48\textwidth]{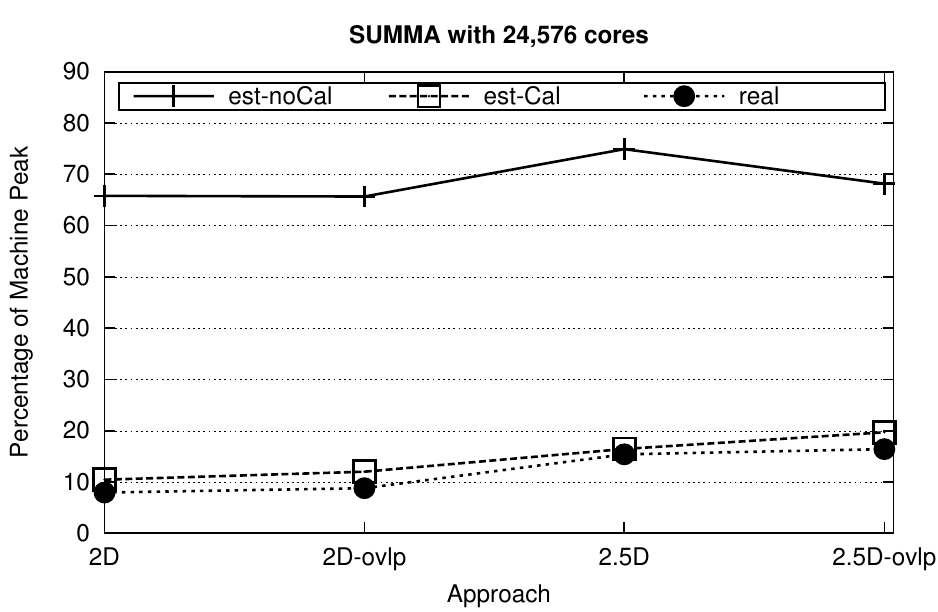}
  \caption{Real and estimated performance for SUMMA using matrices of 32,768$\times$32,768 doubles}
  \label{fig:summa}
\end{figure*}

Figures~\ref{fig:cannon} and~\ref{fig:summa} illustrate the comparison for the matrix multiplication algorithms (Cannon's and SUMMA, respectively) using matrices with 32,768$\times$32,768 doubles. It must be remarked that the estimations using the calibration factor allow to rank the different versions of both algorithms correctly according to their performance. The differences with the real results are always less than 4\% of the machine peak for Cannon's and 7\% for SUMMA. 

\begin{figure*}[!ht]
  \centering
  \includegraphics[width=0.48\textwidth]{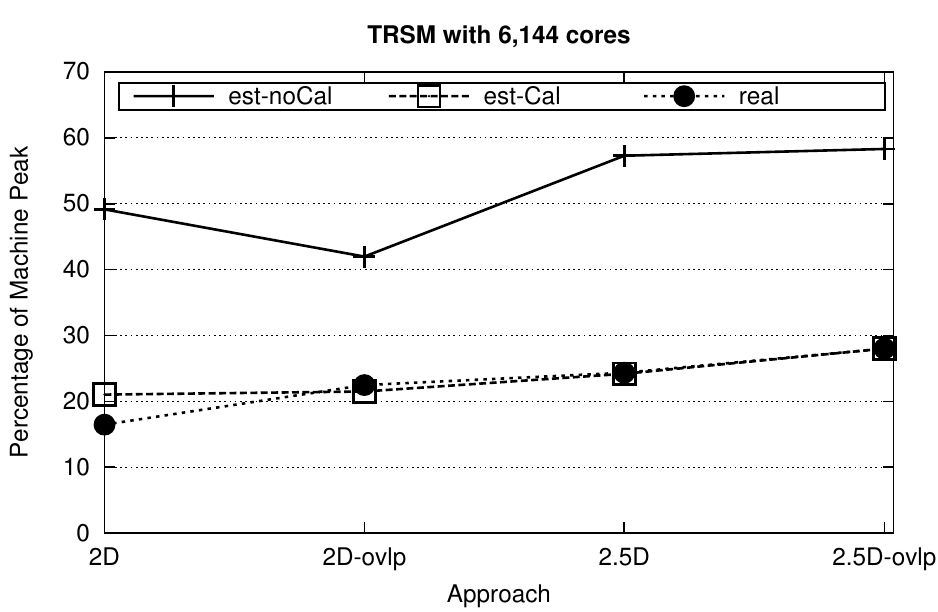}
  \includegraphics[width=0.48\textwidth]{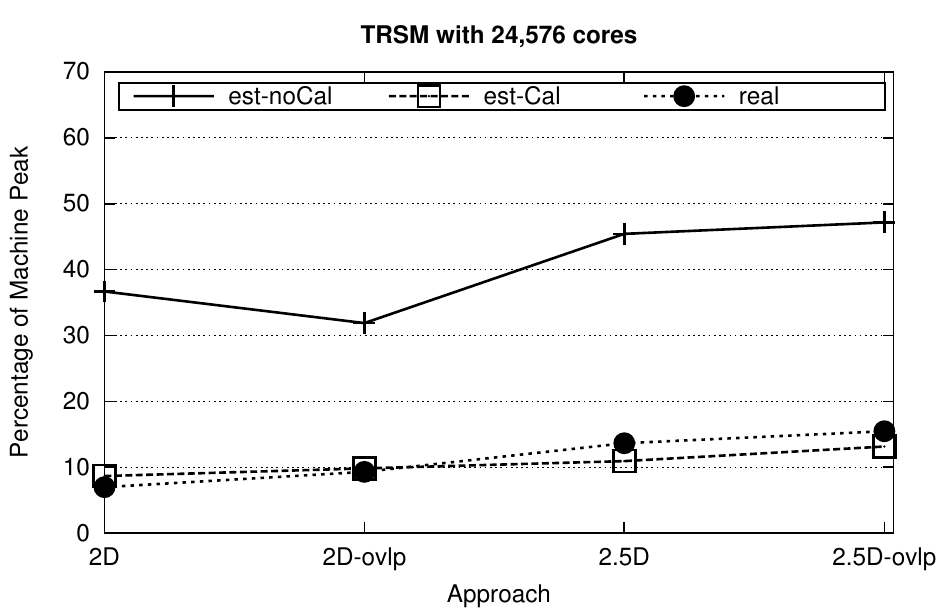}
  \caption{Real and estimated performance for TRSM using matrices of 65,536$\times$65,536 doubles}
  \label{fig:trsm}
\end{figure*}

\begin{figure*}[!ht]
  \centering
  \includegraphics[width=0.48\textwidth]{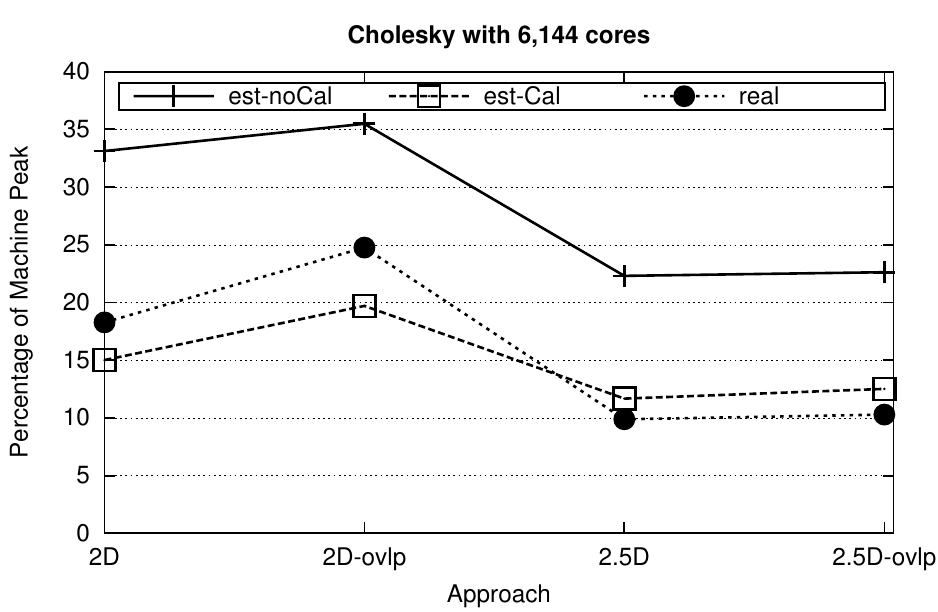}
  \includegraphics[width=0.48\textwidth]{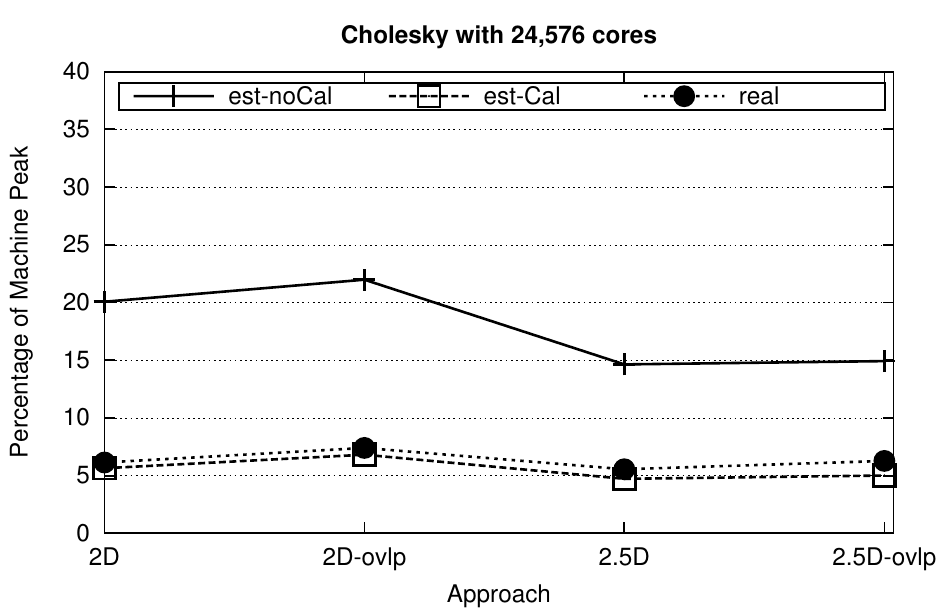}
  \caption{Real and estimated performance for Cholesky using matrices of 65,536$\times$65,536 doubles}
  \label{fig:cholesky}
\end{figure*}

The comparison of the experimental results and estimations for TRSM and Cholesky factorization with matrices of 65,536$\times$65,536 doubles are shown in Figures~\ref{fig:trsm} and~\ref{fig:cholesky}, respectively. They demonstrate that the increase of complexity of the algorithm does not lead to worse estimations.   

\subsection{Predictions for Executions with Additional Resources}
\label{sec:prediction}

As previously mentioned, one of the main advantages of having performance models available is to assist deciding which is the best algorithm for a certain scenario (type of machine, number of cores and problem size). 
%Additionally, the development of the models help the programmers to better understand the behavior of the algorithms and, maybe, improve them. For instance,  communication volume by among rows of the grid is not the same as by columns in the triangular solve and Cholesky. Thus, the knowledge that the calibration factor depends on the distance of the communications can help us to increase performance by ordering the processes in the grid so that the dimension with more communications has the closest processes (and thus less performance degradation). 
Furthermore, programmers can also benefit from performance models to simulate what would happen on scenarios with more resources than the available testbeds. Which algorithm would be the best if infinite memory was available? What would be the trend of the algorithms if the number of cores was increased? These questions can be easily answered with appropriate performance models. 

Tables~\ref{tab:pred-cannon} -~\ref{tab:pred-cholesky} summarize the performance estimations for the four approaches of Cannon's, SUMMA, TRSM and Cholesky factorization. The approach with the best performance for each scenario (number of cores and matrix size) is in boldface. Estimations are obtained for matrices larger than those we could experimentally test and up to 393,216 cores. The maximum calibration factor is the only parameter used in our methodology that depends on the total number of cores. We could not execute the benchmark to measure this factor for more than 24,576 cores. A polynomial regression was employed to estimate the maximum calibration factor for a larger number of cores. 

The behavior of the two matrix multiplication algorithms (Cannon's and SUMMA) is similar. If the workload per core is quite large the 2D approach with overlapping obtains the highest performance. Nevertheless, when the number of cores increases without changing the matrix size the avoidance of communication is more effective. Thus, for a certain matrix size there is a point (in terms of number of cores) where the 2.5D algorithm with overlapping becomes the best choice. This behavior is not reproduced in the TRSM case, where the model predicts that the 2.5D algorithm with overlapping is the best option in all cases. Regarding Cholesky, as in the matrix multiplication algorithms, there is a sweet-spot in terms of number of cores where the 2.5D version with overlapping outperforms the other ones. These results confirm that communication avoiding algorithms combined with overlapping are needed on the way to the exascale era.

\begin{table*}[!ht]
  \centering
  \begin{tabular}{| c || c | c | c | c || c | c | c | c |}
    \hline 
    Size $\rightarrow$ & \multicolumn{4}{ c ||}{32768} & \multicolumn{4}{ c |}{65536}\\
    \hline
    Cores $\downarrow$ & 2D & 2D\_overlp & 2.5D & 2.5D\_overlp & 2D & 2D\_overlp & 2.5D & 2.5D\_overlp \\
    \hline
    1,536 & 67.95 & {\bf 83.69} & 53.63 & 55.56 & 72.36 & {\bf 80.40} & 64.52 & 65.91\\
    \hline
    6,144 & 35.42 & {\bf 59.88} & 35.95 & 37.96 & 50.20 & {\bf 73.20} & 48.22 & 50.95\\
    \hline
    24,576 & 12.87 & 15.33 & 21.56 & {\bf 27.80} & 22.59 & 30.73 & 34.51 & {\bf 45.78}\\
    \hline
    98,304 & 4.57 & 4.93 & 9.37 & {\bf 10.55} & 8.71 & 9.78 & 17.04 & {\bf 21.04}\\
    \hline
    393,216 & 1.30 & 1.35 & 3.94 & {\bf 4.19} & 2.78 & 2.91 & 7.55 & {\bf 8.32}\\
    \hline
  \end{tabular}
  \caption{Predicted percentage of peak flop for the Cannon's algorithm.}
  \label{tab:pred-cannon}
\end{table*}

\begin{table*}[!ht]
  \centering
  \begin{tabular}{| c || c | c | c | c || c | c | c | c |}
    \hline 
    Size $\rightarrow$ & \multicolumn{4}{ c ||}{32768} & \multicolumn{4}{ c |}{65536}\\
    \hline
    Cores $\downarrow$ & 2D & 2D\_overlp & 2.5D & 2.5D\_overlp & 2D & 2D\_overlp & 2.5D & 2.5D\_overlp \\
    \hline
    1,536 & 52.29 & {\bf 68.59} & 49.18  & 46.65 & 62.43 & {\bf 66.47} & 61.19 & 55.19\\
    \hline
    6,144 & 24.98 & 27.85 & 30.28 & {\bf 34.74} & 38.82 & {\bf 58.69} & 43.54 & 43.37\\
    \hline
    24,576 & 10.46 & 12.02 & 16.44 & {\bf 19.71} & 18.92 & 24.28 & 27.67 & {\bf 38.51}\\
    \hline
    98,304 & 4.01 & 4.29 & 7.93 & {\bf 8.75} & 8.75 & 9.83 & 14.68 & {\bf 17.51}\\
    \hline
    393,216 & 1.27 & 1.33 & 3.56 & {\bf 3.77} & 3.62 & 3.84 & 7.75 & {\bf 8.56}\\
    \hline
  \end{tabular}
  \caption{Predicted percentage of peak flop for the SUMMA algorithm.}
  \label{tab:pred-summa}
\end{table*}

\begin{table*}[!ht]
  \centering
  \begin{tabular}{| c || c | c | c | c || c | c | c | c |}
    \hline 
    Size $\rightarrow$ & \multicolumn{4}{ c ||}{65536} & \multicolumn{4}{ c |}{131072}\\
    \hline
    Cores $\downarrow$ & 2D & 2D\_overlp & 2.5D & 2.5D\_overlp & 2D & 2D\_overlp & 2.5D & 2.5D\_overlp \\
    \hline
    1,536 & 43.40 & 39.85 & 41.37 & {\bf 44.16} & 56.10 & 49.62 & 55.58 & {\bf 57.89}\\
    \hline
    6,144 & 21.04 & 21.50 & 24.20 & {\bf 28.00} & 33.49 & 32.39 & 38.01 & {\bf 42.03 }\\
    \hline
    24,576 & 8.70 & 9.84 & 10.94 & {\bf 13.16} & 15.87 & 17.10 & 20.12 & {\bf 26.06}\\
    \hline
    98,304 & 3.33 & 3.60 & 4.42 & {\bf 4.79} & 6.85 & 7.88 & 9.13 & {\bf 10.59}\\
    \hline
    393,216 & 1.24 & 1.29 & 1.38 & {\bf 1.43} & 2.87 & 3.06 & 3.11 & {\bf 3.29}\\
    \hline
  \end{tabular}
  \caption{Predicted percentage of peak flop for the triangular solve (TRSM).}
  \label{tab:pred-trsm}
\end{table*}

\begin{table*}[!ht]
  \centering
  \begin{tabular}{| c || c | c | c | c || c | c | c | c |}
    \hline 
    Size $\rightarrow$ & \multicolumn{4}{ c ||}{65536} & \multicolumn{4}{ c |}{131072}\\
    \hline
    Cores $\downarrow$ & 2D & 2D\_overlp & 2.5D & 2.5D\_overlp & 2D & 2D\_overlp & 2.5D & 2.5D\_overlp \\
    \hline
    1,536 & 32.29 & {\bf 32.29} & 21.02 & 21.81 & 46.88 & {\bf 58.26} &29.86  &  30.72 \\
    \hline
    6,144 & 15.02 & {\bf19.71 } & 11.68 & 12.51 & 18.44 & {\bf26.19 } & 14.78 & 15.96 \\
    \hline
    24,576 & 5.64 & {\bf 6.82} & 4.73 & 5.01 &6.36  & {\bf8.79 } & 6.47 &  6.60\\
    \hline
    98,304 & 1.89 &  {\bf2.01}& 1.83 & 1.87 &4.67  & {\bf5.45} &4.29  & 4.29 \\
    \hline
    393,216 & 0.56 & 0.57 & 0.59 & {\bf0.61 } &  1.66& 1.74 &  1.76& {\bf 1.83} \\
    \hline
  \end{tabular}
  \caption{Predicted percentage of peak flop for Cholesky.}
  \label{tab:pred-cholesky}
\end{table*}

\section{Conclusions}
\label{sec:conclusions}

In this paper we described how to construct performance models for parallel linear algebra algorithms that use communication avoiding and communication overlapping techniques. The proposed methodology was evaluated using two matrix multiplication algorithms (Cannon's and SUMMA), triangular solve (TRSM) and Cholesky factorization. For each of them, we explored different optimization techniques and constructed performance models for every version: communication avoiding, communication overlapping, and both.

The proposed models rely on the use of different parameters measured through portable benchmarks. These parameters are the efficiency of the computational kernels involved in the linear algebra algorithms (calls to multithreaded BLAS routines in our case), the latency and the ideal bandwidth of the network, and calibration factors that encapsulate the communication performance degradation that occurs when several processes concurrently access the network.

For all algorithms in this paper, the performance models can accurately predict the empirical results from the experiments on our target machine, a Cray XE6 system. We have also confirmed that incorporating the communication calibration factor in the models is critical to obtain correct estimations. Moreover, our models are quite flexible because they can take into account runtime constraints (e.g., available memory). These performance models can be useful in a number of applications, such as performance tuning for current systems, performance extrapolation for future systems, and new algorithm design. 

Finally, although the methodology was developed focusing on a specific target architecture, it can be exported to systems with other characteristics as the benchmarks are portable. Proving if this methodology is enough to characterize the performance of dense linear algebra algorithms on other types of large supercomputers or if more parameters must be taken into account is considered future work.

\section*{Acknowledgments}
This research was supported in part by the Office of Science of the U.S. Department of Energy (DE-AC02-05CH11231), DARPA (HR0011-10-9-0008), the Ministry of Science and Innovation of Spain (TIN2010-16735) with FEDER funds of the European Union and the Ministry of Education of Spain under the FPU research grant AP2008-01578. This research used resources of the National Energy Research Scientific Computing Center. We gratefully thank Costin Iancu and Katherine Yelick for their valuable comments to improve this work. 

\bibliographystyle{IEEEtran}
\bibliography{CCGrid14}

\end{document}